\documentclass[iop, tighten, numberedappendix, letterpaper]{emulateapj}

\usepackage{amssymb,amsmath,mathrsfs,graphicx}
\usepackage{verbatim}

\def\h2{$\rm H_2$}

\newcommand{\msun}{M$_{\odot}$}

\newcommand{\lsun}{L$_{\odot}$}
\newcommand{\halpha}{H$\alpha$}

\begin{document}

\shortauthors{Weisz et al.}

\title{The Star Formation History of Leo~T from Hubble Space Telescope Imaging\footnote{Based on observations made with the NASA/ESA Hubble Space Telescope, obtained from the Data Archive at the Space Telescope Science Institute, which is operated by the Association of Universities for Research in Astronomy, Inc., under NASA contract NAS 5-26555}}

\author{Daniel R. Weisz\altaffilmark{1}, 
Daniel B. Zucker\altaffilmark{2,3},
Andrew E. Dolphin\altaffilmark{4},
Nicolas F. Martin\altaffilmark{5},
Jelte T. A. de Jong\altaffilmark{6},
Jon A. Holtzman\altaffilmark{7},
Julianne J. Dalcanton\altaffilmark{1},
Karoline M. Gilbert\altaffilmark{1,10},
Benjamin F. Williams\altaffilmark{1},
Eric F. Bell\altaffilmark{8},
Vasily Belokurov\altaffilmark{9},
N. Wyn Evans\altaffilmark{9}
}

\altaffiltext{1}{University of Washington; dweisz@astro.washington.edu}
\altaffiltext{2}{Macquarie University}
\altaffiltext{3}{Australian Astronomical Observatory}
\altaffiltext{4}{Raytheon}
\altaffiltext{5}{Max-Planck-Institut f\"{u}r Astronomie}
\altaffiltext{6}{Leiden Observatory}
\altaffiltext{7}{New Mexico State University}
\altaffiltext{8}{University of Michigan}
\altaffiltext{9}{University of Cambridge}
\altaffiltext{10}{Hubble Fellow}

\begin{abstract}

We present the star formation history (SFH) of the faintest known star-forming galaxy, Leo~T, based on deep imaging taken with the Hubble Space Telescope (HST) Wide Field Planetary Camera 2 (WFPC2).   The HST/WFPC2 color-magnitude diagram (CMD) of Leo~T  is exquisitely deep, extending $\sim$ 2 magnitudes below the oldest main sequence turnoff, permitting excellent constraints on star formation at all ages.  We use a maximum likelihood CMD fitting technique to measure the SFH of Leo~T assuming three different sets of stellar evolution models:  Padova (solar-scaled metallicity) and BaSTI  (both solar-scaled and $\alpha$-enhanced metallicities).  The resulting SFHs are remarkably consistent at all ages, indicating that our derived SFH is robust to the choice of stellar evolution model.  From the lifetime SFH of Leo~T, we find that 50\% of the total stellar mass formed prior to z $\sim$ 1 (7.6 Gyr ago).  Subsequent to this epoch, the SFH of Leo~T is roughly constant until the most recent $\sim$ 25 Myr, where the SFH shows an abrupt drop.  This decrease could be due to a cessation of star formation or stellar initial mass function sampling effects, but we are unable to distinguish between the two scenarios. Overall, our measured SFH is consistent with previously derived SFHs of Leo~T.  However, the HST-based solution provides improved age resolution and reduced uncertainties at all epochs.  The SFH, baryonic gas fraction, and location of Leo~T are unlike any of the other recently discovered faint dwarf galaxies in the Local Group, and instead bear strong resemblance to gas-rich dwarf galaxies (irregular or transition), suggesting that gas-rich dwarf galaxies may share common modes of star formation over a large range of stellar mass ($\sim$ 10$^{5}-10^{9}$ \msun).

\end{abstract}

\keywords{
galaxies: individual (Leo T dIrr); galaxies: stellar content; Local Group
}

\section{Introduction}
\label{intro}

The discovery of dozens of faint dwarf galaxies in the Local Group has extended our studies of galaxy formation and evolution to the lowest end of the galaxy luminosity function \citep[e.g.,][]{zuc04, wil05, bel07, mar09}.  With extremely low masses and luminosities, the discovery of these galaxies has facilitated new insight into key topics such as the nature of dark matter profiles, stellar feedback and chemical evolution, and the so-called `missing satellites' problem in the framework of CDM cosmologies \citep[e.g.,][]{moo99, wal09, gov10, pen10}.

Leo~T remains unique among the known extremely faint galaxy population.  It has a low luminosity \citep[$M_{V} \sim$ $-$8;][]{irw07, dej08b} and high dark matter content \citep[\msun/\lsun $\sim$ 60-160;][]{rya08, sim07}, yet, unlike other recent discoveries, Leo~T is located in relative isolation \citep[420 kpc from the Milky Way][]{irw07, dej08b}, exhibits evidence for multiple generations of star formation \citep[][]{irw07, dej08a,dej08b}, and has a baryonic gas fraction \citep[$M_{gas}$/($M_{star} + M_{gas}$) $\sim$ 0.8;][]{rya08} comparable to other nearby dwarf irregular galaxies \citep[e.g.,][]{wei11a}.  The mixture of these unusual characteristics provides a truly extreme and unusual environment in which we can study the effects of star formation and stellar feedback in relation to theories of low mass galaxy evolution \citep[e.g.,][]{dek86, orb08, ric09}.

In this paper, we present the color-magnitude diagram (CMD) and the star formation history (SFH) of Leo~T based on imaging obtained with the Wide Field Planetary Camera 2 \citep[WFPC2;][]{hol95} aboard the Hubble Space Telescope (HST).  The long integration times of HST/WFPC2 observations make for an exquisitely deep CMD that extends below the ancient main sequence turnoff (MSTO; $\gtrsim$ 10 Gyr),  allowing for well-constrained measures of star formation at all ages.

This paper is organized as follows. In \S \ref{obs} we detail the WFPC2 observations and photometric reductions and we present the resultant CMD in \S \ref{cmd}.  We then describe our method for measuring the SFH, including the use of multiple stellar evolution models, in \S \ref{sfh}.  Finally, we present and analyze the lifetime cumulative and absolute SFHs of Leo~T in \S \ref{lifetime}.  The conversion between age and redshift used in this paper assume a standard WMAP-7 cosmology as detailed in \citet{jar11}.

\section{Observations and Photometry}
\label{obs}

Observations of a single central field in Leo~T were taken with WFPC2/HST from 21 October 2007 to 29 October 2007 as part of HST program GO-11084 (PI: D.~Zucker).  The set of observations consists of 16 F606W (wide V) and 26 F814W (I) images with total integration times of 19200s and 31200s, respectively.  The 2.5\arcmin $\times$ 2.5\arcmin\ WFPC2 field encloses most of the area defined by the half-light radius \citep[$r_{h} =$ 1.5\arcmin;][]{mar08}.  

We performed PSF photometry on each of the images using HSTPHOT, a stellar photometry package designed for use with WFPC2 \citep{dol00}.  We culled the catalog of detected objects to only include well-measured stars by applying the following photometric criteria: $SNR_{F606W}$ $>$ 4 and $SNR_{F814W}$ $>$ 4 and ($sharp_{F606W}$ $+$ $sharp_{F814W}$)$^{2}$ $\le$ 0.075, yielding 3847 well-measured stars. Definitions of the photometric quality metrics can be found in \citet{dol00} and \citet{dal09}.  The photometric catalog of well-measured stars is available as a high level science product via the HST archive.\footnote[1]{http://archive.stsci.edu/hlsp}

To characterize completeness and observational uncertainties we conducted 500,000 artificial star tests.  After applying the photometric quality criteria to the recovered artificial stars, we measured the 50\% completeness limits to be $m_{F606W} =$ 27.5 and $m_{F814W} =$ 26.9.

\section{The Color Magnitude Diagram}
\label{cmd}

\begin{figure*}[th]
\begin{center}
\epsscale{1.0}
\plotone{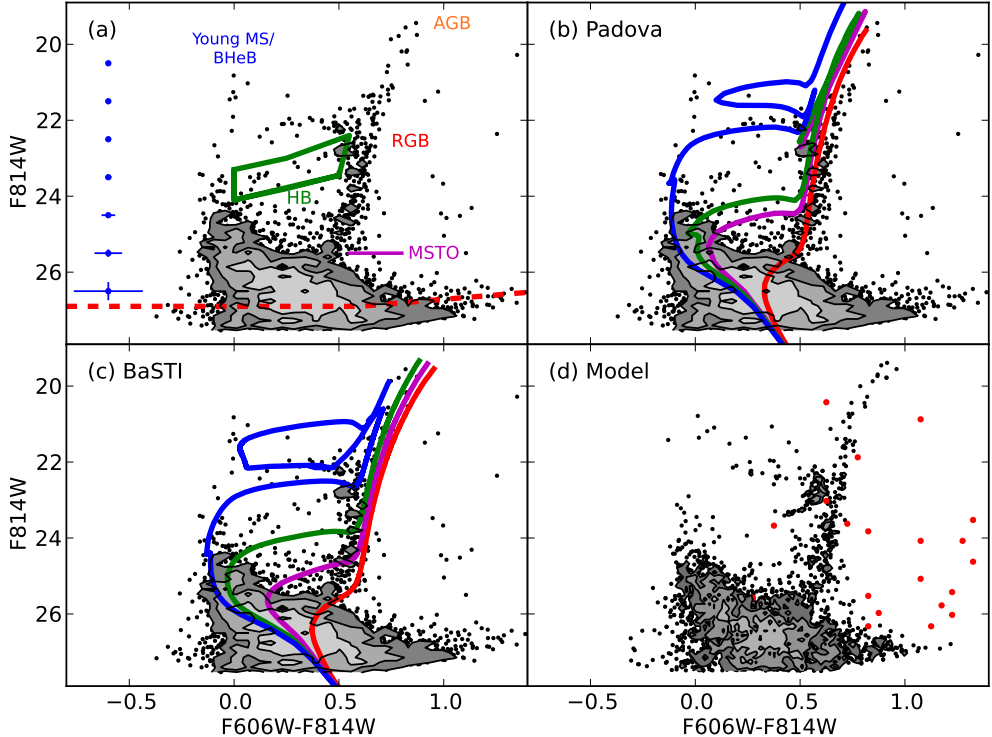}
\caption{\small The deep HST/WFPC2 CMD of Leo~T, corrected  for foreground reddening using the values from \citet{sch98}. \textit{Panel (a):} The red-dashed line represents the 50\% completeness limit and the blue error bars indicate photometric uncertainties, both determined by artificial star tests.  We have also highlighted several age sensitive features, including identification of the horizontal branch (green box) and oldest MSTO (magenta). \textit{Panel (b):} The observed CMD of Leo~T with select Padova isochrones of 500 Myr (blue), 2 Gyr (green), 5 Gyr (magenta), and 10 Gyr (red) and a metallicity of $Z =$ 0.0005, the value derived from the SFH measurement.  \textit{Panel (c):} Same as panel (b) only with the BaSTI isochrones, and a metallicity of $Z =$ 0.0003.  The slight difference in metallicities is due to available values on the respective web interfaces (Padova: http://stev.oapd.inaf.it/cgi-bin/cmd; BaSTI: http://albione.oa-teramo.inaf.it/).  These two panels illustrate some of the metallicity dependent differences in evolved star models, as discussed more extensively in \citet{gal05}.  \textit{Panel (d):} A model CMD from the most likely SFH based on the Padova stellar libraries.  The red points are an example of the stars used to model the intervening foreground population.}
\label{leotcmd}
\end{center}
\end{figure*}

As highlighted in Figure \ref{leotcmd}, the CMD of Leo~T has many interesting characteristics that suggest multiple episodes of star formation.  First, there are several indicators of ancient star formation, such as the oldest MSTO (F606W-F814W $\sim$ 0.5; F814W $\sim$ 26), blue and red horizontal branch (HB) populations (defined by the green box in Figure \ref{leotcmd}), and the red giant branch (RGB; extending vertically between F814W $\sim$ 19 and 25.5).  The broadness of this latter sequence suggests the presence of multiple age and/or metallicity populations.  Second, we see a sparse sampling of stars that are fainter than the HB (between F814W $\sim$ 24.5 and 25.3 and F606W-F814W $\sim$ 0.2 to 0.4).  These stars are consistent with magnitudes and colors of intermediate age ($\sim$ 2-10 Gyr ago) MSTO stars.  Third, the population of stars located near the bright limit of the RGB (F814W $\sim$ 20) may be luminous intermediate age asymptotic giant branch stars (AGBs).  However, given the low number of stars in the CMD, it is difficult to visually discern whether these are truly AGB stars or are associated with the tip of the RGB population.  Finally, Leo~T has a small number of luminous MS and blue helium burning stars (F814W $\sim$ 21-22, F606W-F814W $\sim$ 0-0.3), which both trace star formation within the most recent $\sim$ 1 Gyr.  The lack of extremely luminous MS stars suggests that Leo~T has had little or no star formation at very recent times.

\section{Measuring the Star Formation History}
\label{sfh}

We have measured the SFH of Leo~T using the CMD fitting package MATCH \citep{dol02}.  Briefly, MATCH builds sets of synthetic CMDs from user defined  parameters including a stellar initial mass function (IMF), binary fraction, a searchable range of distance and extinction values, and fixed bin sizes in age, metallicity, color and magnitude, and then convolves the model CMD with observational biases as measured from artificial star tests.  MATCH then employs a maximum likelihood statistic to compare synthetic and observed CMDs.  The SFH that corresponds to the best matched synthetic CMD is the most probable SFH of the observed population. A full description of MATCH can be found in \citet{dol02}.

To measure the SFH of Leo~T, we selected a Kroupa IMF \citep{kro01} with a mass range of 0.15 to 120 \msun\ and a binary fraction of 0.35, where the mass of the secondary is drawn from a uniform mass distribution.  We selected the solar-scaled metallicity Padova set of stellar evolution models with updated low mass AGB track \citep[e.g.,][]{gir10} for our primary SFH measurement.  We also solved for  SFHs of Leo~T using the BaSTI stellar evolution libraries with both solar-scaled and $\alpha$-enhanced models \citep{pie04}.  These particular models provide age sensitivity from $\lesssim$ 25 Myr ago to 14 Gyr ago, which is needed to accurately model the mixed-age population of Leo~T.  With all models, we designated a search range of metallicities of [M/H] = $-$2.3 to $-$1.2, with resolution of 0.1 dex.  The lower metallicity limit is set by the availability of the models. We initially explored a higher metallicity upper limit, but found this space to be largely unexplored by the code.  

For the Padova models, we defined 40 logarithmic time bins over the range $\log(t) =$ 6.6-10.15.  The bins were spaced by 0.05 dex for $\log(t) =$ 9.0-10.15 and 0.1 dex for $\log(t) =$ 7.4-9.0.  Given the lack of extremely luminous stars on the MS, we designated a single time bin for the youngest ages, $\log(t) =$ 6.6-7.4.  For solutions using the BaSTI models, we designated 39 identical time bins to the Padova scheme, but excluded the youngest time bins as the BaSTI models only extend to $\log(t) =$ 7.4 and 7.45 for the solar-scaled and $\alpha$-enhanced models, respectively.  We did not attempt solutions using other models such as Dartmouth \citep[][]{dot08} because they do not cover the full range in ages spanned by stars in Leo~T.  To facilitate comparison between the models and observations, the observed CMD and synthetic CMDs were binned with a resolution of 0.1 in magnitude and 0.05 in color.

Well-known differences between the Padova and BaSTI HB models can introduce significant biases into the measured SFHs \citep[e.g.,][]{gal05}.  We have mitigated these potential biases by placing the HB into a single Hess diagram bin, whose dimensions are indicated by the green box in Figure \ref{leotcmd}.  In effect, this process requires that each model generate a HB, but the precise details of the HB population (e.g., luminosity and morphology) do not strongly affect the measured SFH.  The most secure leverage on the ancient SFH comes from the oldest MSTO.

Additionally, to account for intervening Milky Way foreground populations, we used the  `foreground' utility included in the MATCH package to construct model foreground CMDs that were used in the derivation of the SFH.  This utility produces a CMD-based on the results of \citet{dej10}, who measured thick disk and halo structural parameters using MATCH.  The model foreground CMDs were made with the Dartmouth stellar evolution models \citep{dot08} that include stars with masses as low as 0.1 \msun.  Given the small amount of expected foreground contamination in the Leo~T CMD, we do not anticipate that the particular choice of models will substantially influence the measured SFH.  However, the Dartmouth models provide a more complete census of the intervening low mass Galactic stellar population, resulting in a more accurate accounting of any foreground contamination.

We quantify uncertainties in the SFHs using a set of Monte Carlo tests.  The Monte Carlo tests are designed to account for uncertainties due to the number of stars on the CMD (random uncertainties) and biases due to uncertainties in the stellar models (systematic uncertainties).  To quantify the random uncertainties we sample the best-fit CMD using  a Poisson random noise generator.  We then introduce additive errors in $M_{bol}$ and $\log(T_{eff})$ and resolve for the SFH.  These additive values serve as a proxy for systematic uncertainties in the underlying stellar models by mimicking the scatter in SFH uncertainties obtained by using alternate isochrone sets \citep[e.g., BaSTI, Dartmouth;][]{pie04, dot08} to fit the data. The SFH of the new CMD is then measured identically to the original solution, constituting a single Monte Carlo realization.  We found that the uncertainties were stable after 50 Monte Carlo tests, and thus conducted 50 realizations. A more detailed description of this process can be found in \citet{wei11a}.

For each set of models, we allowed MATCH to determine the best combination of SFR, metallicity, distance, and extinction.  For each of the three solutions, MATCH found a best fit extinction corrected distance modulus of 23.05 and foreground extinction of $A_{V} =$ 0.20.  These values favorably compare with previously derived distances of 23.10$\pm$0.2 \citep{irw07, dej08b} and the foreground extinction value of $A_{V} =$ 0.1 from \citet{sch98}.  We show the simulated CMD from the best fit SFH from the Padova solution in panel (d) of Figure \ref{leotcmd}.

\section{The Lifetime Star Formation History of Leo~T}
\label{lifetime}

\begin{figure}
\begin{center}
\epsscale{1.2}
\plotone{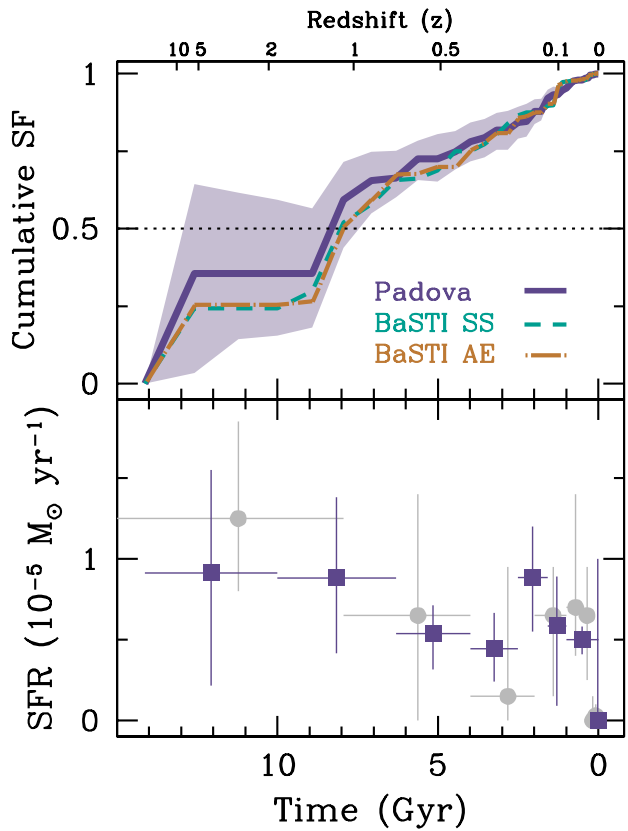}
\caption{{\it Top panel} -- The lifetime cumulative SFH, i.e., the fraction of total stellar mass formed prior to a given epoch, of Leo~T measured using the Padova solar-scaled metallicity stellar evolution models (purple) and the BaSTI models (solar-scaled: green; $\alpha$-enhanced: orange). The lightly shaded purple region represents the 1-$\sigma$ uncertainties on the SFH measured with the Padova models. {\it Bottom panel} -- The lifetime absolute SFH of Leo~T, as measured with the Padova models (purple squares).  The plotted uncertainties in the y-direction reflect the 1-$\sigma$ uncertainties on the SFR, which those in the x-direction indicate the width of the time bin. For comparison, we have over-plotted the SFH measured by \citet{dej08b} as grey circles. }
\label{leotsfh}
\end{center}
\end{figure}

In Table \ref{tab1} and Figure \ref{leotsfh}, we present the cumulative and absolute SFHs of Leo~T.  The cumulative SFH, i.e., the fraction of the total stellar mass formed at a given time, provides a normalized measure of the stellar mass accumulation in Leo~T.   Compared to the absolute SFHs,  cumulative SFHs are less affected by correlated SFRs in adjacent time bins, allowing us to plot the cumulative SFHs at full time resolution.  Regions where the cumulative SFH exhibits zero growth for extended periods or has large uncertainties indicate intervals over which our knowledge of the shape of the absolute SFH is uncertain.  We therefore utilize the cumulative SFH to inform appropriate time binning for the absolute SFH. For a more detailed discussion of optimizing time resolution for CMD-based SFHs, see Appendix A in \citet{wei11a}.

The best fit SFHs from the Padova (purple) and BaSTI models ($\alpha$-enhanced: orange; solar-scaled: green) show excellent consistency.  Specifically, considering the mean cumulative SFHs for each model, we see that Leo~T formed 50\% of its total stellar mass prior to z $\sim$ 1 (7.6 Gyr ago).  For ages younger than this, Leo~T exhibits a nearly constant SFH.  For times prior to z $\sim$ 1 (7.6 Gyr ago), the amplitude of the uncertainty envelope indicates that we cannot place tight constraints on the precise epochs of star formation (i.e., we cannot reliably distinguish between separate bursts or extended constant star formation).  The similarity in the SFHs derived with different models suggests that the primary cause of the large uncertainties is the small number of truly ancient stars in the CMD.  On the whole, we find that over the WFPC2 field of view, the total stellar mass formed in Leo~T is 1.05$_{-0.23}^{+0.27}$$\times$10$^{5}$ \msun.  This derived value is comparable to the present day stellar mass estimate of $1.2$$\times$10$^{5}$ \msun\ derived by \citet{rya08}.

In the bottom panel of Figure \ref{leotsfh}, we have plotted the absolute SFH derived with the Padova model as purple squares, and have over-plotted the SFH based on imaging taken with the Large Binocular Telescope \citep[LBT;][]{dej08b} as grey circles.  Comparing the two solutions, we see only subtle differences.  First, due to the significantly deeper observations, the HST-based SFH affords higher time resolution while maintaining SFR uncertainties that are comparable to or smaller in amplitude than those presented in \citet{dej08b}.  Specifically, we have binned the HST-based solution with a time resolution of $\Delta \log(t) =$ 0.2, compared to a value 0.3 for the LBT solution.  Along with the higher time resolution, we also see a decrease in SFR uncertainties.  Both of these improvements are particularly evident at epochs more recent than z $\sim$ 0.5 ($\sim$ 5 Gyr ago), where our solution confirms that Leo~T had a nearly constant SFH to a high degree of confidence.  Within the most recent 1 Gyr, the small number of luminous MS and blue core helium burning stars result in large fractional uncertainties on the SFH, restricting our ability to decipher precise patterns of star formation. 

Within the past 25 Myr, the precise SFR of Leo~T is difficult to quantify.  On one hand, there are no luminous MS or core helium burning stars, which can be interpreted as a lack of recent star formation.  However, the effects of stochastic IMF sampling also provide an alternative explanation \citep[e.g.,][]{das11, fum11}.  In this scenario, it is possible that the recent SFR of Leo~T is sufficiently low ($\lesssim$ 10$^{-5}$ \msun/yr) that star formation is continuous, but no massive stars are actually formed.  Unfortunately, SFH and IMF effects are largely degenerate \citep[e.g.,][]{mil79, elm06}, and we are not able to distinguish between the two scenarios.  As a result, we can only conclude that the upper limit on the recent SFR in Leo~T is $\sim$ 10$^{-5}$ \msun--a SFR for which stochastic IMF sampling would not produce any luminous, young stars.  Integrated tracers of recent star formation in Leo~T reveal only faint GALEX ultra-violet fluxes and no \halpha\ emission \citep{lee11, ken08}, which could be consistent with truncated star formation or a stochastically sampled IMF scenario. 

For epochs prior to z $\sim$ 2 (10 Gyr ago), we find marginal improvement over the LBT solution.  The HST-based SFH confirms the relatively high SFR at ancient times seen in the LBT solutions.   On the whole, we conclude that the LBT and HST SFHs are in excellent agreement.

The CMD fitting process additionally provides a rough estimate for the metallicity of Leo~T.  For solutions derived from each stellar library, we find a mean isochronal metallicity of [M/H] $\sim$ $-$1.6 -- $-$1.8, which does not exhibit significant variance over the lifetime of Leo~T.  Uncertainties on the mean metallicity are $\sim$ 0.3 dex.  The mean isochronal metallicities are slightly more metal rich than the mean spectroscopic value of [Fe/H] $\sim$ $-$2.3$\pm$0.1, with a spread of 0.35 dex \citep{sim07}.  A direct comparison between isochronal and spectroscopically derived metallicities is challenging due to issues such as RGB star selection effects and the conversion from a canonical metallicity, i.e., [M/H], to [Fe/H] \citep[e.g.,][]{lia11}.  Due to these uncertainties, we are only able to state that the isochronal metallicities from the derived SFHs are coarsely comparable to the spectroscopically determined values.

\section{Summary and Conclusions}

We present the lifetime SFH of the Local Group gas-rich, faint dwarf galaxy Leo~T based on deep HST/WFPC2 imaging.  The HST imaging covers nearly the entire area defined by the half-light radius, and the resulting CMD extends $\sim$ 2 mag below the ancient MSTO.  Using the MATCH CMD fitting routine, we measured three SFHs of Leo~T using the Padova (solar-scaled) and BaSTI (solar-scaled and $\alpha$-enhanced) stellar evolution models.  For all models considered, we found virtually identical SFHs, confirming the robustness of measurement.  The SFH of Leo~T shows that 50\% of its total stellar mass was formed prior to z $\sim$ 1 (7.6 Gyr ago) and that the SFH at times younger than this epoch is approximately constant.  The sparse sampling of young stars makes the shape of the SFH within the most recent 1 Gyr uncertain.  The SFH of Leo~T in the past $\sim$ 25 Myr is uncertain due to the lack of luminous MS and  BHeB stars.  Their absence could either be due to a truncation in star formation or stochastic sampling effects of the IMF, but we cannot distinguish between the two scenarios.  We also found little evolution in the isochronal metallicities of Leo~T, such that over the coarse of its lifetime it is consistent with a constant value of $[M/H] \sim$ $-$1.6.

\begin{figure}[h]
\begin{center}
\epsscale{1.2}
\plotone{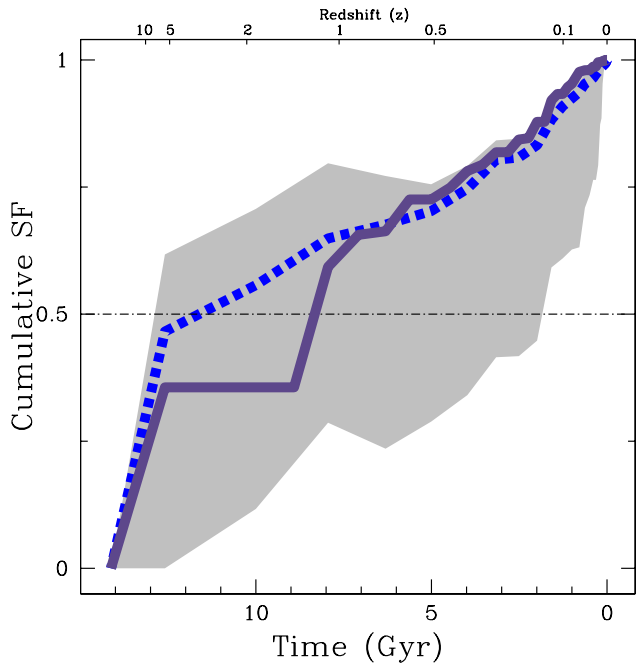}
\caption{A comparison between the average cumulative SFH of $\sim$ 30 nearby dwarf irregular galaxies (blue dashed line) as presented in \citet{wei11a} and that of Leo~T (solid purple line).   The grey shaded region is the uncertainty in the mean cumulative SFH of the larger dwarf irregular sample.  At nearly all times, the SFH of Leo~T is in good agreement with that of a typical dIrr.  The apparent discrepancy between the two SFHs from $\sim$ 8 -12 Gyr is largely a visual effect.  This interval coincides with large uncertainties on the ancient SFH of Leo~T, and the two solutions are consistent within both sets of uncertainties. The overall similarity in the two SFHs provides further evidence that Leo~T is an extremely low mass dwarf irregular galaxy.}
\label{leotdirr}
\end{center}
\end{figure}

The SFH of Leo~T provides additional insight into its true morphological type.  Although Leo~T has a similar stellar and dark matter mass to other recently discovered faint Local Group dwarf galaxies, its SFH, gas content, and location bear strong resemblance to nearby dwarf irregular galaxies  \citep[e.g.,][see Figure \ref{leotdirr};]{mat98, dol05, tol09, wei11a,wei11b}.  The added information from the SFH reinforces previous suggestions that Leo~T may be a Phoenix-like transition dwarf galaxy \citep[e.g.,][]{irw07, rya08}, i.e., gas-rich with current SF \citep[e.g.,][]{mat98}.  However, extensive discussion of transition dwarfs in \citet{wei11a}, suggests that galaxies with this designation are actually dwarf irregular galaxies that are either in between episodes of star formation or in the process of permanently losing gas due to an external disturbance.  Coupling this discussion with its empirical characteristics Leo~T appears to be the lowest mass dwarf irregular galaxy known to date.  Consequently, this finding suggests that the gas-rich dwarf galaxies share common star formation processes across the entire known dwarf galaxy mass spectrum ($\sim 10^{5} - 10^{9}$ \msun).

\section*{Acknowledgements}

DRW would like to thank Jorge Penarrubia for insightful discussions on extremely low mass galaxies.  NFM acknowledges funding by Sonderforschungsbereich SFB 881 "The Milky Way System" (subproject A3) of the German Research Foundation (DFG). Support for KMG is provided by NASA through Hubble Fellowship grant HST-HF-51273.01 awarded by the Space Telescope Science Institute. This work is based on observations made with the NASA/ESA Hubble Space Telescope, obtained from the data archive at the Space Telescope Science Institute.  Partial support for this work was provided by NASA through grants GO-11084 and AR-10945 from the Space Telescope Science Institute, which is operated by AURA, Inc., under NASA contract NAS5-26555. This research has made use of the NASA/IPAC Extragalactic Database (NED), which is operated by the Jet Propulsion Laboratory, California Institute of Technology, under contract with the National Aeronautics and Space Administration.  This research has made extensive use of NASA's Astrophysics Data System Bibliographic Services.

\begin{deluxetable}{ccccc}
\small
\tablecolumns{5}
\tablehead{
\colhead{Youngest Bin Time} &
\colhead{Oldest Bin Time} &
\colhead{Absolute SFR} &
\colhead{Cumulative SF Fraction} &
\colhead{}  \\
\colhead{$\log(t)$} &
\colhead{$\log(t)$} &
\colhead{(10$^{-6}$ \msun\ yr$^{-1}$)} &
\colhead{} &
\colhead{} \\
\colhead{(1)} &
\colhead{(2)} &
\colhead{(3)} &
\colhead{(4)} &
\colhead{} 
}

\tablecaption{The Star Formation History of Leo~T}
\startdata
\hline
 & & Padova Model & &\\
 \hline
       6.60 & 9.00 & 5.02$_{-0.930}^{+0.800}$ & 1.00$_{-0.00}^{+0.00}$ &\\
       9.00 & 9.20 & 5.87$_{-4.95}^{+3.05}$ & 0.95$_{-0.01}^{+0.01}$ &\\
       9.20 & 9.40 & 8.85$_{-3.33}^{+3.16}$ & 0.92$_{-0.02}^{+0.03}$ &\\
       9.40 & 9.60 & 4.46$_{-2.04}^{+2.22}$ & 0.84$_{-0.05}^{+0.05}$ &\\
       9.60 & 9.80 & 5.39$_{-2.23}^{+1.74}$ & 0.78$_{-0.06}^{+0.06}$ &\\
       9.80 & 10.00 & 8.83$_{-4.67}^{+4.98}$ & 0.66$_{-0.06}^{+0.09}$ &\\
       10.00 & 10.15 & 9.15$_{-6.97}^{+6.37}$ & 0.36$_{-0.20}^{+0.23}$ &\\
       \hline
       & & BaSTI Solar-Scaled & &\\
       \hline
       7.40 & 9.00 & 2.38$_{-1.86}^{+1.98}$ & 1.00$_{-0.00}^{+0.00}$ &\\
       9.00 & 9.20 & 12.7$_{-4.06}^{+5.01}$ & 0.98$_{-0.03}^{+0.02}$ &\\
       9.20 & 9.40 & 3.07$_{-2.34}^{+2.37}$ & 0.90$_{-0.07}^{+0.05}$ &\\
       9.40 & 9.60 & 7.40$_{-1.83}^{+2.38}$ & 0.87$_{-0.09}^{+0.06}$ &\\
       9.60 & 9.80 & 3.76$_{-3.76}^{+5.70}$ & 0.75$_{-0.11}^{+0.09}$ &\\
       9.80 & 10.00 & 10.1$_{-3.06}^{+2.56}$ & 0.66$_{-0.18}^{+0.14}$ &\\
       10.00 & 10.15 & 5.51$_{-5.51}^{+9.24}$ & 0.24$_{-0.19}^{+0.31}$ &\\
        \hline
       & & BaSTI $\alpha$-enhanced & &\\
       \hline
       7.45 & 9.00 & 2.51$_{-1.52}^{+1.40}$ & 1.00$_{-0.00}^{+0.00}$ &\\
       9.00 & 9.20 & 11.2$_{-4.07}^{+4.00}$ & 0.97$_{-0.01}^{+0.01}$ &\\
       9.20 & 9.40 & 4.62$_{-2.63}^{+2.35}$ & 0.90$_{-0.05}^{+0.04}$ &\\
       9.40 & 9.60 & 6.56$_{-1.90}^{+1.74}$ & 0.85$_{-0.05}^{+0.05}$ &\\
       9.60 & 9.80 & 3.09$_{-3.09}^{+6.35}$ & 0.75$_{-0.06}^{+0.06}$ &\\
       9.80 & 10.00 & 10.1$_{-2.80}^{+3.17}$ & 0.67$_{-0.13}^{+0.11}$ &\\
       10.00 & 10.15 & 5.68$_{-5.68}^{+6.86}$ & 0.25$_{-0.24}^{+0.27}$ 
       
\enddata
\tablecomments{\scriptsize The absolute and cumulative SFHs of Leo~T as derived with the Padova and BaSTI stellar evolution models.  Columns (1) and (2) indicate the youngest and oldest ages of the respective time bins.  The absolute SFR in column (3) reflects the SFR over the duration of the time bin.  The cumulative SF value is the fraction of total stellar mass formed prior to the time indicated in column (1).  The listed uncertainties represent the 16th and 84th percentiles measured from the set of Monte Carlo realizations (\S \ref{sfh}). By construction the cumulative SFH is zero at $\log(t) =$ 10.15.}
\label{tab1}
\end{deluxetable}

\end{document}